# Real-time Video Processing in Web Applications

**Cristian Ioniţă**
Academy of Economic studies
Bucharest, Romania
crionita@ie.ase.ro

**Alexandru Bărbulescu**
Academy of Economic studies
Bucharest, Romania
alexbarbulescu@ie.ase.ro

**ABSTRACT**

The OpenGL ES standard is implemented in modern desktop and mobile browsers through the WebGL API. This paper explores the potential for using OpenGL ES hardware acceleration for real time video processing in standard HTML5 applications. It analyses the WebGL performance across device types and compares it with the standard JavaScript and canvas performance.

**ACM Classification Keywords**

H.5.2 User Interfaces

**INTRODUCTION**

The current paper proposes a method of using hardware acceleration available in current devices in order to enable real time video analysis and processing in standard web applications. It measures the performance of the proposed solution, compares it with existing techniques and determines the range of devices and algorithm types that are feasible using the current methods.

**HARDWARE ACCELERATION IN WEB APPLICATIONS**

WebGL (Web Graphics Library) is a JavaScript API (Application Programming Interface) which offers access to 2D and 3D rendering in any compatible web browser. One of its advantages is that no add-on or plugin is necessary. This is used in HTML <canvas> elements. WebGL was initially developed by Vladimir Vukićević ([1]) working for Mozilla Foundation [3]. Initial release was in March 2011 but a stable release has been made two years later in March 2013. It is cross platform library also based on OpenGL ES [4]. The first experiments of the author were made in 2006 and in 2007 Mozilla and Opera had made their individual experiments in order to prove the usability of the concept. In 2009 the Kronos Group consortium a group, WebGL Working Group where the major web browsers developers were invited after 2013 it was widely adopted as a standard.

Although the standard was adopted in 2013 the technology was adopted by the industry in the last year. The rapid adoption in the last 12 months (May 2014 – May 2015) confirms the industry interest in the technology across all device types: desktop 62% → 90% and mobile 29% → 78%.

One important concept is that WebGL allow using low level programming and the power of the hardware graphics processing unit. The library is working with two buffers: frontbuffer which is the image currently visible and backbuffer which is the image being rendered. By programming we see only one image. There is transparent for programmers when the backbuffer is moved to the frontbuffer. The browser can move the backbuffer any time to the frontbuffer except during the execution of the JavaScript code.

WebGL can draw points and lines but the basic shape for 2D and 3D drawing is the triangle. This is a primitive figure for 3D drawings because a plane is uniquely defined by 3 points. WebGL is using the parallel power of the GPU (graphics processing unit) and the integrated or shared memory to compute the final image.

In the last ten years the evolution of the computer graphic cards transformed them from a component that renders images processed by CPU to one that delivers graphics processing units being able to make real time processing and transformation. In programming concepts, these moved from Fixed Function Pipeline (FFP) to Programmable Pipeline (or shaders) as in [2]. In a programmable pipeline with GPU the programmer establish the vertex transformation and fragment processing using a high level programming languages like GLSL. The compiled programs are loaded to the GPU and the task are automatically executed. This is made by a complex system consisting of hardware and drivers. The drivers are exposing the hardware functionalities to the programmers and graphics libraries.

The modern GPU exposes fully programmable hardware units also named shaders units. The name came from the idea that the units are connected in a pipeline and the output of one shading unit is the input for the next. The Vertex Shader is the programmable Shader stage in the rendering pipeline that handles the processing of individual vertices. According to OpenGL documentation Vertex shaders are fed Vertex Attribute data, as specified from a vertex array object by a drawing command. A vertex shader receives a single vertex from the vertex stream and generates a single vertex to the output vertex stream. A Geometry Shader (GS) is a Shader program written in GLSL that governs the processing of Primitives. Geometry shaders reside between the Vertex Shaders (or the optional Tessellation stage) and the fixed-function Vertex Post-Processing stage. A geometry shader is optional and does not have to be used. After creation of vertices geometry we need to create pixels. This is made by the rasterization process. This process take all primitives and split into individual fragment which is colored by fragment shader and is turned into frame buffer. The programmable fragment shader unit takes the fragments produced by the rasterization process and executes an algorithm provided by a graphics developer to produce the final color, depth and stencil values for each fragment. This part can be used to achieve special visual effects, including post-processing filters.

One special part of the image processing is the use of parallelization. Processing the images involve many repetitive operations that are very time consuming. The architecture of modern GPUs is based on many parallel execution units and is appropriate for data parallel algorithms.

**KERNEL IMAGE PROCESSING**

The bitmap image is represented by a matrix o pixels. Many image processing algorithms can be expressed using a convolution process to apply a kernel to an image. This is made by applying a mask also named filter matrix or kernel. It involves making matrix operations in order to calculate the result matrix. In the result matrix every pixel value is calculated from the initial matrix multiplied with the kernel matrix. In most of cases the kernel matrix consist in 5x5 or 3x3 values. These are enough for obtaining the most of the effects.

| 10 | 52 | 63 | 42 | 74 |
|----|----|----|----|----|
| 86 | 24 | 45 | 28 | 82 |
| 62 | 91 | 17 | 24 | 2  |
| 49 | 19 | 18 | 36 | 75 |
| 41 | 15 | 78 | 17 | 14 |

| 0 | 1 | 0 |
|---|---|---|
| 0 | 0 | 1 |
| 0 | 0 | 0 |

**Figure 1. Convolution matrix**

The new value for 45 is calculated in the following manner:

52x0+63x1+42x0+24x0+45x0+28x1+91x0+17x0x24x0=91

The method can be used for implementing edge detection algorithms such as Sobel [5] and Frei-Chen [6]. The goal of an edge detection algorithm is to identify points of an image where the intensity changes abruptly. There are many factors that need to be taken into consideration. There is surface orientation, discontinuities, lightning changes, very similar textures, etc. The aim of edge detection is to apply edge detectors to an image and to receive a set of lines or curves that delimits objects. The results is a gray-scale image where each pixel value indicates whether it is or not on the boundary of an object. There are many algorithms and the results are very sensitive to image characteristics. Usually it's better to use multiple edge detection algorithms and choose the one which is best for the case. In every algorithm there is a needed to establish the threshold. For every pixel if the value is above the threshold it is considered part of the edge, otherwise not. After this step named binarization we can use other algorithms to discover the edges in the image.

The Sobel filter uses two 3x3 convolution matrices to detect vertical and horizontal gradients of the image.

$$G_x = \begin{bmatrix} -1 & -2 & -1 \\ 0 & 0 & 0 \\ 1 & 2 & 1 \end{bmatrix}, \quad G_y = \begin{bmatrix} -1 & 0 & 1 \\ -2 & 0 & 2 \\ -1 & 0 & 1 \end{bmatrix},$$

$$|\Delta I| = \sqrt{(G_x * I)^2 + (G_y * I)^2}$$

**Figure 2. Sobel edge detection matrix.**

These masks are applied to the 3x3 footprint of every RGB color in the image. The results are then used to obtain the gradient value.

Another algorithm used for edge detection is Frei-Chen edge detector. The algorithm uses nine 3x3 convolution masks. The weighted sum of all convolution results is used to determine the final value for each pixel.

$$G_1 = \frac{1}{2\sqrt{2}}\begin{bmatrix} 1 & \sqrt{2} & 1 \\ 0 & 0 & 0 \\ -1 & -\sqrt{2} & -1 \end{bmatrix} \quad G_2 = \frac{1}{2\sqrt{2}}\begin{bmatrix} 1 & 0 & -1 \\ \sqrt{2} & 0 & -\sqrt{2} \\ 1 & 0 & -1 \end{bmatrix} \quad G_3 = \frac{1}{2\sqrt{2}}\begin{bmatrix} 0 & -1 & \sqrt{2} \\ 1 & 0 & -1 \\ -\sqrt{2} & 1 & 0 \end{bmatrix}$$

$$G_4 = \frac{1}{2\sqrt{2}}\begin{bmatrix} \sqrt{2} & -1 & 0 \\ -1 & 0 & 1 \\ 0 & 1 & -\sqrt{2} \end{bmatrix} \quad G_5 = \frac{1}{2}\begin{bmatrix} 0 & 1 & 0 \\ -1 & 0 & -1 \\ 0 & 1 & 0 \end{bmatrix} \quad G_6 = \frac{1}{2}\begin{bmatrix} -1 & 0 & 1 \\ 0 & 0 & 0 \\ 1 & 0 & -1 \end{bmatrix}$$

$$G_7 = \frac{1}{6}\begin{bmatrix} 1 & -2 & 1 \\ -2 & 4 & -2 \\ 1 & -2 & 1 \end{bmatrix} \quad G_8 = \frac{1}{6}\begin{bmatrix} -2 & 1 & -2 \\ 1 & 4 & 1 \\ -2 & 1 & -2 \end{bmatrix} \quad G_9 = \frac{1}{3}\begin{bmatrix} 1 & 1 & 1 \\ 1 & 1 & 1 \\ 1 & 1 & 1 \end{bmatrix}$$

**Figure 3. Frei-Chen edge detection matrices. [6]**

The first four kernels are used for edges, the next four are used for determining lines and the last one is used for smoothing out the result. The projection equation used by the algorithm is presented in figure 4.

$$\cos e = \sqrt{\frac{M}{S}} \quad \text{where } M = \sum_{k \in \{e\}} (G_k * I)^2 \text{ and } S = \sum_{k=1}^{9} (G_k * I)^2$$

**Figure 4. Frei-Chen projection equation. [6]**

From practice the results of Frei-Chen algorithm are better because the algorithm is less sensitive to noise and is able to detect edges with small gradients. Also Sobel can be improved by a normalization factor which in Frei-Chen is represented by the ninth mask [6].

**CANVAS IMAGE PROCESSING**

Using HTML5 canvas element to render the processed video image involves two steps.

In the initialization step the canvas element is created and added to the DOM tree. The canvas has the same size as the source video. The context object is also created in this step and saved for using in the display loop.

The second step is setting the display loop. The display loop is based on the *requestAnimationFrame* method. For each animation frame the current video frame is drawn on the canvas. Before being displayed, the image is processed by applying one or more convolution operators. Processing is performed on the *ImageData* objects obtained by using the *getImageData* method of the *canvas context* object.

## GLSL PROCESSING

The proposed WebGL solution uses the OpenGL ES rendering pipeline to process the source video frames. Figure 5 shows the solution architecture.

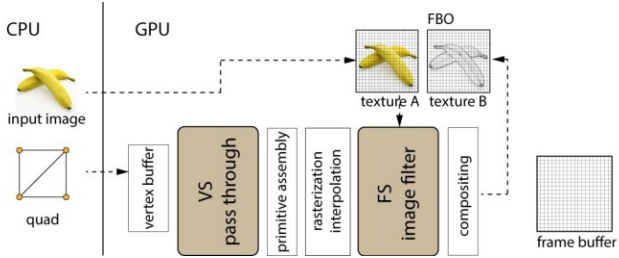

**Figure 5. OpenGL pipeline for image processing [7]**

In the WebGL initialization phase we create the GLSL program and the necessary attributes and uniforms (video frame texture, kernels …). The program consists of two shaders. The first one is a simple pass-through vertex shader that processes the quad used to display the processed frame. The second one is the fragment shader that applies the convolution operators for every pixel of the texture. The simplified single matrix variant:

```
precision mediump float;
uniform sampler2D u_image;
uniform vec2 u_textureSize;
uniform float u_kernel[9];
varying vec2 v_texCoord;
void main() {
vec2 onePixel = vec2(1.0, 1.0) / u_textureSize;
vec4 colorSum = texture2D(u_image, v_texCoord +
onePixel * vec2(-1, -1)) * u_kernel[0] + ...;
float kernelWeight = u_kernel[0] + u_kernel[1] +
...;
...
gl_FragColor = vec4((colorSum / kernelWeight).rgb,
1.0); } }
```

The video frame processing loop is constructed using the *requestAnimationFrame* method. For each video frame the following operations are performed: the current video frame is transferred from the video element to the GPU texture memory and the OpenGL ES pipeline is executed in order to process and display the video frame.

## BENCHMARK SETUP

In order to determine the merits and performance characteristics of the two image processing techniques across devices we set up a benchmark.

The hardware used for benchmarking is representative for the desktop, laptop and smartphone devices.

For measurement we used *DOMHighResTimeStamp* objects (accurate to the thousandth of millisecond) obtained through *Performance API*. This technique was used to measure image acquisition and processing times.

|  | Desktop | Laptop | Phone |
|---|---|---|---|
| Description | i7-920, 18GB, HD 7870 XT | i7-4500U, 8GB, HD Graphics 4400 | HTC One M8 - Krait 400, 2GB, Adreno 330 |
| CPU Cores | 4 | 2 | 4 |
| CPU Frequency | 2.67 GHz | 1.8 – 3 GHz | 2.3 GHz |
| GPU Cores | 1536 | 20 | 4 |
| GPU Frequency | 925 – 975 MHz | 200 – 1100 MHz | 550 MHz |

**Table 1. Hardware used for benchmarking.**

The actual WebGL drawing / processing time can't be measured directly from the browser. Because of this we used and indirect method based on the fact that the browser will honor a *requestAnimationFrame* request only after the current drawing operation is completed. This technique allows us to obtain an adequate measurement for the WebGL performance for values >16ms.

## PERFORMANCE ANALYSIS

### Real time webcam edge detection

The first test was performed using the webcam as a video source. The video was obtained using the *getUserMedia* method. The resulting 640x480 video stream was channeled to a visible *video* element on the page and used as source for WebGL and *canvas* processing.

|  | Desktop | Laptop | Phone |
|---|---|---|---|
| **WebGL** FPS | 60 | 60 | 34 |
| Acq. / Proc. | 0.4 / * | 1.3 / * | 5.2 / * |
| **canvas** FPS | 37 | 32 | 7.2 |
| Acq. / Proc. | 0.4 / 24.6 | 1.1 / 26.6 | 4.8 / 111 |

**Table 2. Sobel edge detection on 640x480 webcam video. Acquisition and processing times in milliseconds.**

From the results presented in table 1 we can see that even applying a relatively simple convolution operator in real time is possible only with the more powerful x86 CPUs. Only those CPUs have the necessary speed to apply the operators in the required 30ms time frame using a single thread.

Using the GPU shaders for applying the operator in real time is possible an all devices. Even the low power Adreno 330 is able to finish the task in under 30ms.

The image acquisition time (transfer from the video stream to canvas or GPU texture memory) is similar for *WebGL* and *canvas* across devices.

In the following two subsections we analyze the most important factors that influence the performance for this class of problems: video size (total number of pixels that need processing) and number of operations per pixel.

**Sensitivity to video source size**

In order to determine the sensitivity of the processing in respect to the number of processed pixels we use the same video re-encoded at 4 standard resolutions. The video was streamed from the web server and the processing was performed in real time on the client.

|  | **Desktop** | **Laptop** | **Phone** |
|---|---|---|---|
| 320x240 (0.1 mp) | 60 / 60 | 35 / 60 | 16 / 50 |
| 640x480 (0.3 mp) | 30 / 60 | 19 / 60 | 7 / 45 |
| 1280x720 (0.9 mp) | 16 / 60 | 10 / 58 | 3 / 40 |
| 1920x1080 (2.1 mp) | 7 / 60 | 4 / 40 | 1.5 / 35 |

**Table 3. Average number of frames per second using *canvas / WebGL*.**

Unlike the canvas processing time which evolves linearly, the WebGL processing time remains almost constant. Almost all FPS differences in the WebGL case are caused by variations in HTTP transfer, video decoding and frame drawing time.

**Sensitivity to processing complexity**

All the tests performed up to now used the relatively simple Sobel operator (2 matrices). Other operators (like the Frei-Chen presented above – 9 matrices) and filter combinations can require a larger number of operations. In order to determine the canvas / WebGL sensitivity to the number of operations performed per pixels we apply the same convolution matrix multiple times per pixel.

| **Operators** | **Desktop** | **Laptop** | **Phone** |
|---|---|---|---|
| 2 | 60 | 60 | 53 |
| 10 | 60 | 60 | 48 |
| 20 | 60 | 60 | 43 |
| 50 | 60 | 60 | 55 |
| 100 | 60 | 60 | 20 |
| 500 | 60 | 51 | 3.5 |

**Table 4. Average number of frames per second using *WebGL*.**

The results show that the WebGL processing technique presented above can be used to apply at least 50 convolution operator even on a mobile phone. The desktop computer was able to process up to 25000 matrix operations maintaining a frame rate above 30 FPS. The canvas results are not presented in the table because the performance degraded sharply after 2 operations.

**FUTURE RESEARCH**

Because the results of the proposed WebGL processing technique are very promising we plan to extend our research on more complex image processing and computer vision algorithms. The next two types of problems we plan to address are motion detection (control UI using webcam) and object recognition (especially 1D and 2D barcode scanning).

Another future direction research is combining the presented GLSL processing technique with the new Web Workers API for exploiting the multiple CPU cores available in modern devices for algorithms that can't be parallelized efficiently on the GPU.

**CONCLUSION**

In this paper we presented an improved method of processing images in the context of HTML applications. Our measurements show that *WebGL* much better suited for real time video processing even on mobile devices. The standard canvas-based processing is not fast enough for most video processing tasks. In the future the performance gap between the two techniques will probably increase even more bases on the fact that the number GPU cores available in mobile devices will grow much faster than the number of cores and the performance improvements in WebGL 2 / OpenGL ES 3.0.